\documentclass[12pt]{article}
\usepackage{amssymb}

\begin{document}

\newcommand{\sect}[1]{\setcounter{equation}{0}\section{#1}}
\renewcommand{\theequation}{\thesection.\arabic{equation}}
\newcommand{\be}{\begin{equation}}
\newcommand{\ee}{\end{equation}}
\newcommand{\bea}{\begin{eqnarray}}
\newcommand{\eea}{\end{eqnarray}}
\newcommand{\nonu}{\nonumber\\}
\newcommand{\beano}{\begin{eqnarray*}}
\newcommand{\eeano}{\end{eqnarray*}}
\newcommand{\eps}{\epsilon}
\newcommand{\om}{\omega}
\newcommand{\vph}{\varphi}
\newcommand{\sig}{\sigma}
\newcommand{\CC}{\mbox{${\mathbb C}$}}
\newcommand{\RR}{\mbox{${\mathbb R}$}}
\newcommand{\QQ}{\mbox{${\mathbb Q}$}}
\newcommand{\ZZ}{\mbox{${\mathbb Z}$}}
\newcommand{\NN}{\mbox{${\mathbb N}$}}
\newcommand{\1}{\mbox{\hspace{.0em}1\hspace{-.24em}I}}
\newcommand{\II}{\mbox{${\mathbb I}$}}
\newcommand{\prt}{\partial}
\newcommand{\und}[1]{\underline{#1}}
\newcommand{\wh}[1]{\widehat{#1}}
\newcommand{\wt}[1]{\widetilde{#1}}
\newcommand{\mb}[1]{\ \mbox{\ #1\ }\ }
\newcommand{\half}{\frac{1}{2}}
\newcommand{\noin}{\not\!\in}
\newcommand{\topa}[2]{\genfrac{}{}{0pt}{}{#1}{#2}}
\newcommand{\rhotimes}{\mbox{\raisebox{-1.2ex}{$\stackrel{\displaystyle\otimes}
{\mbox{\scriptsize{$\rho$}}}$}}}
\newcommand{\bin}[2]{{\left( {#1 \atop #2} \right)}}
\newcommand{\A}{{\cal A}}
\newcommand{\B}{{\cal B}}
\newcommand{\C}{{\cal C}}
\newcommand{\F}{{\cal F}}
\newcommand{\R}{{\cal R}}
\newcommand{\T}{{\cal T}}
\newcommand{\W}{{\cal W}}
\newcommand{\cS}{{\cal S}}
\newcommand{\bS}{{\bf S}}
\newcommand{\cL}{{\cal L}}
\newcommand{\hlp}{{\RR}_+}
\newcommand{\hlm}{{\RR}_-}
\newcommand{\Hil}{{\cal H}}
\newcommand{\D}{{\cal D}}
\newcommand{\bg}{{\bf g}}
\newcommand{\alg}{\C_{\cS}}
\newcommand{\balg}{\B_{\cS}}
\newcommand{\frep}{\lambda_{\R,\T}}
\newcommand{\srep}{{\widetilde \F}(\C_\cS)}
\newcommand{\rep}{\F(\C_\cS)}
\newcommand{\form}{\langle \, \cdot \, , \, \cdot \, \rangle }
\newcommand{\e}{{\rm e}}
\newcommand{\LL}{\mbox{${\mathbb L}$}}
\newcommand{\Rp}{{R^+_{\, \, \, \, }}}
\newcommand{\Rm}{{R^-_{\, \, \, \, }}}
\newcommand{\Rpm}{{R^\pm_{\, \, \, \, }}}
\newcommand{\Tp}{{T^+_{\, \, \, \, }}}
\newcommand{\Tm}{{T^-_{\, \, \, \, }}}
\newcommand{\Tpm}{{T^\pm_{\, \, \, \, }}}
\newcommand{\baral}{\bar{\alpha}}
\newcommand{\barbt}{\bar{\beta}}
\newcommand{\supp}{{\rm supp}\, }
\newcommand{\blambda}{{\overline{\lambda}}}
\newcommand{\EE}{\mbox{${\mathbb E}$}}
\newcommand{\JJ}{\mbox{${\mathbb J}$}}
\newcommand{\MM}{\mbox{${\mathbb M}$}}
\newcommand{\ct}{{\cal T}}

\newtheorem{theo}{Theorem}[section]
\newtheorem{coro}[theo]{Corollary}
\newtheorem{prop}[theo]{Proposition}
\newtheorem{defi}[theo]{Definition}
\newtheorem{conj}[theo]{Conjecture}
\newtheorem{lem}[theo]{Lemma}
\newcommand{\prf}{\underline{\it Proof.}\ }
\newcommand{\finprf}{\null \hfill {\rule{5pt}{5pt}}\\[2.1ex]\indent}
\newcommand{\ie}{{\it i.e.}\ }

\pagestyle{empty}
\rightline{April 2004}

\vfill

\begin{center}
{\Large\bf The quantum non-linear Schr\"odinger model\\[1.2ex]
with point-like defect}
\\[2.1em]

{\large
V. Caudrelier$^{a}$\footnote{caudreli@lapp.in2p3.fr},
M. Mintchev$^{b}$\footnote{mintchev@df.unipi.it}
and E. Ragoucy$^{a}$\footnote{ragoucy@lapp.in2p3.fr}}\\
\end{center}

\null

\noindent
{\it $^a$ LAPTH, 9, Chemin de Bellevue, BP 110,
F-74941 Annecy-le-Vieux
         cedex, France\\[2.1ex]
$^b$ INFN and Dipartimento di Fisica, Universit\'a di
         Pisa, Via Buonarroti 2, 56127 Pisa, Italy}
\vfill

\begin{abstract}

We establish a family of point--like impurities which preserve the quantum
integrability of the non--linear Schr\"odinger model in 1+1 space--time
dimensions. We briefly describe the construction of the exact second quantized
solution of this model in terms of an appropriate
reflection--transmission algebra.
The basic physical properties of the solution, including the
space--time symmetry
of the bulk scattering matrix, are also discussed.

\end{abstract}

\vfill
\rightline{LAPTH-1037/04}
\rightline{IFUP-TH 15/2004}
\rightline{\tt hep-th/0404144}
\newpage
\pagestyle{plain}
\setcounter{page}{1}


\sect{Preliminaries}

We present in this Letter the exact solution of the quantum non-linear
Schr\"odinger (NLS) model with point-like impurity in 1+1 space-time
dimensions. We focus mainly on the physical properties of the
solution, referring for the mathematical details and proofs to
\cite{CMR}. Being the first exactly solvable example
with non--trivial bulk scattering matrix, the NLS model provides valuable
information about the interplay between point--like impurities,
integrability and symmetries.

Assuming that the impurity is localized at $x=0$, the model we are
concerned with
is defined by the equation of motion
\be
(i\partial_t+\partial_x^2)\Phi(t,x)-2g|\Phi(t,x)|^2\Phi(t,x) = 0\, ,
\qquad x\not= 0\, ,
\label{eqm}
\ee
and the impurity boundary conditions
\be
\left(\begin{array}{cc} \Phi (t,+0) \\ \prt_x \Phi
(t,+0)\end{array}\right) = \alpha
\left(\begin{array}{cc} a & b\\ c&d\end{array}\right)
\left(\begin{array}{cc} \Phi (t,-0) \\ \prt_x \Phi
(t,-0)\end{array}\right) \, ,
\label{bc}
\ee
where
\be
\{a,...,d \in \RR,\, \alpha \in \CC\, :\, ad -bc = 1,\,
{\overline \alpha} \alpha = 1 \} \, .
\label{parameters}
\ee
Eq. (\ref{bc}) captures the interaction of the field $\Phi$ with the impurity
\cite{Coutinho:1999xj,SCH} and deserves some explanation.
The parameters (\ref{parameters}) label the self--adjoint
extensions of the operator $-\prt_x^2$,
defined on the space $C_0^\infty (\RR \setminus \{0\})$ of smooth
functions with compact
support separated  from the origin $x=0$. This operator is not
self--adjoint, but its closure
admits self--adjoint extensions, which are parametrized \cite{A1}
in terms of (\ref{parameters}). In order to avoid the presence of bound states,
we take below $g>0$ and restrict further the parameters (\ref{parameters})
according to:
\be
\left\{\begin{array}{cc}
a+d+\sqrt{(a-d)^2+4} \leq 0 \, ,
& \quad \mbox{$b<0$}\, ,\\[1ex]
c (a+d)^{-1} \geq 0\, ,
& \quad \mbox{$b=0$}\, ,\\[1ex]
a+d-\sqrt{(a-d)^2+4} \geq 0 \, ,
& \quad \mbox{$b>0$}\, . \\[1ex]
\end{array} \right.
\label{nobs}
\end{equation}
The operator $-\prt_x^2$ has no bound states in the domain (\ref{nobs}). A
complete orthonormal system of scattering states is given by
\be
\psi_k^+(x) = \theta(-x) T_-^+(k) \e^{ikx} +
\theta(x)\left [\e^{ikx} + R_+^+(-k)\e^{-ikx}\right ]\, , \quad k<0\, ,
\label{basis1}
\ee
\be
\psi_k^-(x) = \theta(x) T_+^-(k) \e^{ikx} +
\theta(-x)\left [\e^{ikx} + R_-^-(-k)\e^{-ikx}\right ]\, , \quad k>0\, ,
\label{s}
\ee
where $\theta$ denotes the standard Heaviside function and
\bea
R_+^+(k) = \frac{bk^2 + i(a-d)k + c}{bk^2 + i(a+d)k - c} \, , \qquad
T_+^-(k) = \frac{2i\alpha k}{bk^2 + i(a+d)k - c}\, ,
\label{coef1} \\
R_-^-(k) = \frac{bk^2 + i(a-d)k + c}{bk^2 - i(a+d)k - c} \, , \qquad
T_-^+(k) = \frac{-2i{\overline \alpha} k}{bk^2 - i(a+d)k - c}\, ,
\label{coef2}
\eea
are the {\it reflection} and {\it transmission coefficients} from the
impurity.
It is easily verified that the {\it reflection} and {\it transmission
matrices},
defined by
\be
\R(k) = \left(\begin{array}{cc} R_+^+(k) & 0\\0 &
R_-^-(k)\end{array}\right)\, ,
\qquad
\T(k) =\left(\begin{array}{cc} 0 & T_+^-(k)\\T_-^+(k) & 0\end{array}\right)\, ,
\label{rtmat}
\ee
satisfy hermitian analyticity
\be
\R(k)^\dagger = \R(-k)\,  , \qquad  \T(k)^\dagger = \T(k)\,  ,
\label{hanal}
\ee
and unitarity
\bea
\T(k) \T(k) + \R(k) \R(-k)  = \II \, , \\
\T(k) \R(k) +  \R(k) \T(-k) = 0 \,  .
\label{unit}
\eea
Let us observe in passing that the reflection $x\mapsto -x$
leaves invariant eq. (\ref{eqm}), but not always (\ref{bc}).
The parity preserving impurities are selected by
\be
a=d\, , \qquad \alpha = \overline \alpha \, .
\label{parity}
\ee

We conclude this section by pointing out that the
impurity boundary conditions (\ref{bc}) can be implemented, coupling the
field $\Phi$ to an external potential with support in $x=0$. The set
\be
\{a=d=1,\, b=0,\, c=2\eta;\, \alpha=1\}
\label{deltaimp}
\ee
{}for instance, corresponds to the potential
\be
V(x) = 2\eta \delta (x)\, ,
\label{potential}
\ee
known as $\delta$--impurity. A general potential, which incorporates all
four real parameters (\ref{parameters}), has been suggested recently 
in \cite{Langmann}.

\sect{The solution}

When considered on the whole line $\RR$, eq. (\ref{eqm}) defines
one of the most extensively studied integrable systems, which has
been solved \cite{Sklyanin:ye}--\cite{Davies:gc} by means of the
inverse scattering transform \cite{FT}. We will show below that
this method can be extended to eqs. (\ref{eqm},\ref{bc}) as well.
For this purpose we will generalize to the case with impurity the
Rosales \cite{R,Fokas:1995rj} series expansion of the solution in
terms of the scattering data. A similar generalization has already
been used for solving \cite{Gattobigio:1998hn, Gattobigio:1998si}
the boundary value problem associated with (\ref{eqm}) on the the
half--line $\RR_+$.

It is instructive to display first the classical solution of
eqs. (\ref{eqm},\ref{bc}). We introduce the fields $\Phi_\pm$ defined by
\be
\Phi (t, x) =
\left\{\begin{array}{cc}
\Phi_- (t, x) \, ,
& \quad \mbox{$x<0$}\, ,\\[1ex]
\Phi_+ (t, x)\, ,
& \quad \mbox{$x>0$}\, ,\\[1ex]
\end{array} \right.
\label{pm}
\end{equation}
and inspired by \cite{R,Fokas:1995rj} consider the series representation
\be
\Phi_\pm (t, x) =  \sum_{n=0}^\infty (-g)^n\, \Phi^{(n)}_\pm (t,x) \, ,
\label{phiser}
\ee
with
\be
\Phi^{(n)}_\pm (t,x) =
\int\prod_{i=1\atop j=0}^n
\frac{dp_i}{2\pi}\frac{dq_j}{2\pi}\,
\blambda_\pm (p_1)\ldots\blambda_\pm
(p_n)\lambda_\pm(q_n)\ldots\lambda_\pm(q_0)
\frac{e^{i\sum\limits_{j=0}^n(q_j x-q^2_j t)-i
\sum\limits_{i=1}^n(p_i x-p^2_i t)}}
{\prod\limits_{i=1}^n(p_i-q_{i-1})
(p_i-q_i)} \, ,
\label{class}
\ee
where the bar denotes complex conjugation and $\lambda_\pm$ define
two solutions
\be
\Phi_\pm^{(0)}(t,x) = \int \frac{dq}{2\pi} \lambda_\pm (q) \e^{i(qx -
q^2t)} \, ,
\ee
of the free Schr\"odinger equation. For
sufficiently smooth $\lambda_\pm$ the series (\ref{phiser}) converges and
Rosales argument guarantees that $\Phi$ is a solution of (\ref{eqm}).
In order to satisfy the boundary condition (\ref{bc}), we take $\lambda_\pm$ of
the form
\be
\left(\begin{array}{cc} \lambda_+(k) \\ \lambda_-(k)
\end{array}\right) =
\left(\begin{array}{cc} 1 & T_+^-(k)\\
T_+^-(k)&1 \end{array}\right)
\left(\begin{array}{cc} \mu_+(k) \\
\mu_-(k)\end{array}\right) +
\left(\begin{array}{cc} R_+^+(k) & 0\\
0&R_-^-(k) \end{array}\right)
\left(\begin{array}{cc} \mu_+(-k) \\
\mu_-(-k)\end{array}\right)\, ,
\ee
where $\mu_\pm$ are smooth functions with certain decay and 
analyticity properties.
Then the conditions (\ref{nobs}) guarantee the smoothness of $\lambda_\pm$
and using unitarity (\ref{unit}), one easily verifies that
$\lambda_\pm$  satisfy
\begin{eqnarray}
\label{RT1}
\lambda_+(k)=T_+^-(k)~\lambda_-(k)+R_+^+(k)~\lambda_+(-k)\, ,\\
\label{RT2}
\lambda_-(k)=T_-^+(k)~\lambda_+(k)+R_-^-(k)~\lambda_-(-k)\, .
\end{eqnarray}
{}Following \cite{CMR}, one can prove now that
the boundary condition (\ref{bc}) holds order by order in $g$.
The order $n=0$ is a direct consequence of
(\ref{RT1},\ref{RT2}). For checking the higher orders it is convenient
to introduce the new variables $\beta_\pm$ defined by
$$
\left(\begin{array}{cc} \beta_+(k)
\\ \beta_-(k)
\end{array}\right) =\left(\begin{array}{cc} \lambda_+(k)
\\ \lambda_-(k)
\end{array}\right)+
\left(\begin{array}{cc} 0 & \alpha(d+ibk)\\
-{\overline \alpha}(d+ibk)&0 \end{array}\right)
\left(\begin{array}{cc} \lambda_+(-k) \\
\lambda_-(-k)\end{array}\right) \, ,
$$
and use that
$$
\beta_+(k) = -\frac{bk^2-i(a+d)k-c}{bk^2+i(a+d)k-c}\, \beta_+(-k)\ ,
\qquad \beta_-(k) = -\beta_-(-k)\, .
$$
The freedom remaining in the choice of $\mu_\pm$ is fixed by the initial
conditions. We will discuss this point at the quantum level, where the
initial conditions are captured by the canonical commutation relations
(see (\ref{ccr1},\ref{ccr2}) below).

\medskip

We turn now to the quantum case, fixing first of all
the basic structures which are involved in the second
quantization of eqs. (\ref{eqm},\ref{bc}). They are:

\begin{itemize}

\item A Hilbert space $\Hil$ with positive definite scalar product
$\langle \cdot\, ,\, \cdot \rangle$, which describes the states of
the system;

\item An operator valued distribution $\Phi(t,x)$, defined on
a dense domain $\D\subset \Hil$ and satisfying the equation of
motion (\ref{eqm}) and the impurity boundary condition (\ref{bc}) in
mean value on $\D$, as well as the equal time canonical commutation relations
\be
[\Phi(t,x)\, ,\, \Phi(t,y)]= [\Phi^*(t,x)\, ,\, \Phi^*(t,y)]= 0
\, ,
\label{ccr1}
\ee
\be
[\Phi(t,x)\, ,\, \Phi^*(t,y)]= \delta(x-y) \, ,
\label{ccr2}
\ee
where $\Phi^*$ is the Hermitian conjugate of $\Phi $;

\item A distinguished normalizable state $\Omega \in \D$ -- the vacuum,
which is cyclic with respect to the field $\Phi^*$.
\medskip

\end{itemize}

Our goal now is to describe the construction of the elements
$\{\Hil, \D, \Omega, \Phi\}$ with
the above properties. A convenient starting point is the well--known
bulk scattering matrix
\be
S(k_1-k_2) = \frac{k_1-k_2-ig}{k_1-k_2+ig} \, ,
\label{s1}
\ee
of the quantum NLS model without impurity. In terms of (\ref{s1}) we define
the $4\times 4$ matrix
\be
\cS_{\alpha_1 \alpha_2}^{\beta_1 \beta_2}(k_1,k_2) =
S(\alpha_1 k_1 - \alpha_2 k_2)
\delta_{\alpha_1}^{\beta_1}\, \delta_{\alpha_2}^{\beta_2}\, , \qquad
\alpha_i,\, \beta_i = \pm \, ,
\label{bulkS}
\ee
which will turn out to be the bulk scattering matrix with impurity.
As a preliminary step in verifying this statement,
one can show that $\cS$ satisfies unitarity
\be
\cS_{12} (k_1,k_2)\, \cS_{21}(k_2,k_1) = \II\otimes \II \, ,
\label{unit1}
\ee
hermitian analyticity
\be
\cS^\dagger_{12}(k_1,k_2) = \cS_{21}(k_2,k_1) \, ,
\ee
the quantum Yang--Baxter equation
\be
\cS_{12}(k_1,k_2) \cS_{13}(k_1,k_3) \cS_{23} (k_2,k_3)
= \cS_{23}(k_2,k_3) \cS_{13}(k_1,k_3) \cS_{12}(k_1,k_2)  \, ,
\label{qyb}
\ee
and the boundary Yang--Baxter equation
\bea
\cS_{12}(k_1 , k_2)\, \R_1(k_1)\,
\cS_{21}(k_2 , -k_1)\, \R_2(k_2) = \nonumber\\
\R_2(k_2)\, \cS_{12}(k_1 , -k_2)\, \R_1(k_1)\,
\cS_{21}(-k_2 , -k_1)\, ,
\label{SRSR}
\eea
where $\R$ is the reflection matrix (\ref{rtmat}) and
the conventional tensor notation has been used. It is worth stressing that
the entries $\cS_{++}^{++}$ and $\cS_{--}^{--}$ depend on $k_1-k_2$ and are
therefore Galilean invariant. On the contrary, $\cS_{+-}^{+-}$ and
$\cS_{-+}^{-+}$
being functions of $k_1+k_2$ break this symmetry.

The matrix $\cS$ with the properties (\ref{unit1}-\ref{SRSR}) identifies a
reflection--trans\-mission (RT) algebra $\alg$ \cite{Mintchev:2002zd,
Mintchev:2003ue},
which is the basic tool
of our construction. The general concept of RT algebra has been designed for
describing factorized scattering in integrable models with impurities.
In what follows we will show that in the NLS model the algebra $\alg$ allows
to reconstruct the off-shell quantum field $\Phi $ as well. $\alg$ is
an associative
algebra with identity $\bf 1$, particle
$\{a^{\ast \alpha} (k),\, a_\alpha (k)\}$ and
impurity (defect) $\{r_\alpha^\beta (k),\, t_\alpha^\beta (k)\}$
generators obeying:
\medskip

(i) bulk exchange relations
\bea
a_{\alpha_1}(k_1) \, \, a_{\alpha_2 }(k_2) \, \; - \; \,
          \cS_{\alpha_2 \alpha_1 }^{\beta_2 \beta_1 }
          (k_2 , k_1)\,\, a_{\beta_2 }(k_2)\, a_{\beta_1 }(k_1) & = & 0
          \, , \qquad \quad \label{aa}\\
a^{\ast \alpha_1} (k_1)\, a^{\ast \alpha_2 } (k_2) -
          a^{\ast \beta_2 } (k_2)\, a^{\ast \beta_1 } (k_1)\,
          \cS_{\beta_2 \beta_1 }^{\alpha_2 \alpha_1 }(k_2 , k_1) & = & 0
          \, , \qquad \quad \label{a*a*} \\
a_{\alpha_1 }(k_1)\, a^{\ast \alpha_2 } (k_2) \; - \;
          a^{\ast \beta_2 }(k_2)\,
          \cS_{\alpha_1 \beta_2 }^{\beta_1 \alpha_2}(k_1 , k_2)\,
          a_{\beta_1 }(k_1) & = &  \nonumber\\
          2\pi\, \delta (k_1 - k_2)\,
          \left [\delta_{\alpha_1 }^{\alpha_2 }\, {\bf 1} +
t_{\alpha_1}^{\alpha_2}(k_1)\right] +
          2\pi\, \delta (k_1 + k_2)\,  r_{\alpha_1 }^{\alpha_2 }(k_1)
          \, ;
\label{aa*}
\eea

(ii) defect exchange relations
\bea
\cS_{\alpha_1 \alpha_2}^{\gamma_1 \gamma_2}(k_1 , k_2)\,
r_{\gamma_1}^{\delta_1}(k_1)\,
\cS_{\gamma_2 \delta_1}^{\delta_2 \beta_1}(k_2 , -k_1)\,
r_{\delta_2}^{\beta_2}(k_2) = \nonumber\\
r_{\alpha_2}^{\gamma_2}(k_2)\,
\cS_{\alpha_1\gamma_2}^{\delta_1 \delta_2}(k_1 , -k_2)\,
r^{\gamma_1}_{\delta_1}(k_1)\,
\cS_{\delta_2\gamma_1}^{\beta_2 \beta_1}(-k_2 , -k_1)
\, ; \label{rr}
\eea
\bea
\cS_{\alpha_1 \alpha_2}^{\gamma_1 \gamma_2}(k_1 , k_2)\,
t_{\gamma_1}^{\delta_1}(k_1)\,
\cS_{\gamma_2 \delta_1}^{\delta_2 \beta_1}(k_2 , k_1)\,
t_{\delta_2}^{\beta_2}(k_2) = \nonumber\\
t_{\alpha_2}^{\gamma_2}(k_2)\,
\cS_{\alpha_1\gamma_2}^{\delta_1 \delta_2}(k_1 , k_2)\,
t^{\gamma_1}_{\delta_1}(k_1)\,
\cS_{\delta_2\gamma_1}^{\beta_2 \beta_1}(k_2 , k_1)
\, ; \label{tt}
\eea
\bea
\cS_{\alpha_1 \alpha_2}^{\gamma_1 \gamma_2}(k_1 , k_2)\,
t_{\gamma_1}^{\delta_1}(k_1)\,
\cS_{\gamma_2 \delta_1}^{\delta_2 \beta_1}(k_2 , k_1)\,
r_{\delta_2}^{\beta_2}(k_2) = \nonumber\\
r_{\alpha_2}^{\gamma_2}(k_2)\,
\cS_{\alpha_1\gamma_2}^{\delta_1 \delta_2}(k_1 , -k_2)\,
t^{\gamma_1}_{\delta_1}(k_1)\,
\cS_{\delta_2\gamma_1}^{\beta_2 \beta_1}(-k_2 , k_1)
\, ; \label{tr}
\eea

(iii) mixed exchange relations
\be
a_{\alpha_1}(k_1)\, r_{\alpha_2}^{\beta_2}(k_2) =
\cS_{\alpha_2 \alpha_1}^{\gamma_2 \gamma_1}(k_2 , k_1)\,
r_{\gamma_2}^{\delta_2}(k_2)\,
\cS_{\gamma_1 \delta_2}^{\delta_1 \beta_2}(k_1 , -k_2)\,
a_{\delta_1}(k_1)
\, , \qquad \label{ar}
\ee
\be
r_{\alpha_1}^{\beta_1}(k_1)\, a^{\ast \alpha_2}(k_2) =
a^{\ast \delta_2}(k_2)\,
\cS_{\alpha_1 \delta_2}^{\delta_1 \gamma_2}(k_1 , k_2)\,
r_{\delta_1}^{\gamma_1}(k_1)\,
\cS_{\gamma_2\gamma_1}^{\alpha_2 \beta_1}(k_2 , -k_1)\,
\, , \qquad \label{ra*}
\ee
\be
a_{\alpha_1}(k_1)\, t_{\alpha_2}^{\beta_2}(k_2) =
\cS_{\alpha_2 \alpha_1}^{\gamma_2 \gamma_1}(k_2 , k_1)\,
t_{\gamma_2}^{\delta_2}(k_2)\,
\cS_{\gamma_1 \delta_2}^{\delta_1 \beta_2}(k_1 , k_2)\,
a_{\delta_1}(k_1)
\, , \qquad \label{at}
\ee
\be
t_{\alpha_1}^{\beta_1}(k_1)\, a^{\ast \alpha_2}(k_2) =
a^{\ast \delta_2}(k_2)\,
\cS_{\alpha_1 \delta_2}^{\delta_1 \gamma_2}(k_1 , k_2)\,
t_{\delta_1}^{\gamma_1}(k_1)\,
\cS_{\gamma_2\gamma_1}^{\alpha_2 \beta_1}(k_2 , k_1)\,
\, , \qquad \label{ta*}
\ee

(iv) unitarity
\be
t_{\alpha_1}^\beta (k) t^{\alpha_2}_\beta (k) +
r_{\alpha_1}^\beta (k) r^{\alpha_2}_\beta (-k) =
\delta_{\alpha_1}^{\alpha_2} \, ,
\label{unit2}
\ee
\be
t_{\alpha_1}^\beta (k) r^{\alpha_2}_\beta (k) +
r_{\alpha_1}^\beta (k) t^{\alpha_2}_\beta (-k) = 0 \, .
\label{unit3}
\ee

As suggested by (\ref{rtmat}), we assume that $r(k)$ is a diagonal matrix
while $t(k)$ is an anti-diagonal one. Then,
due to the particular form of the $\cal S$--matrix, the
defect relations (ii) are equivalent to
$$
[r_{\alpha_1}^{\beta_1}(k_{1}),r_{\alpha_2}^{\beta_2}(k_{2})]=0\, ,\quad
[r_{\alpha_1}^{\beta_1}(k_{1}),t_{\alpha_2}^{\beta_2}(k_{2})]=0\, ,\quad
[t_{\alpha_1}^{\beta_1}(k_{1}),t_{\alpha_2}^{\beta_2}(k_{2})]=0\, .
$$

The Fock representations $\rep$ of $\alg$ have been classified and explicitly
constructed in \cite{Mintchev:2003ue}. As usual, each Fock representation
involves a cyclic (vacuum) state $\Omega$  obeying
\be
a_\pm (k) \, \Omega = 0 \, .
\label{annihilate}
\ee
We recall also that each $\frep \in \rep$ is uniquely defined by
the doublet $\{\R,\, \T\}$, satisfying eqs.
(\ref{hanal}--\ref{unit},\ref{SRSR}). The quantum version of eqs. 
(\ref{RT1},\ref{RT2}) is
\bea
a_\alpha(k) &=& t_\alpha^\beta (k) a_\beta (k) + r_\alpha^\beta (k) 
a_\beta (-k) \, ,
\label{n1} \\
a^{\ast \alpha}(k) &=& a^{\ast \beta}(k) t_\beta^\alpha (k) +
a^{\ast \beta}(-k) r_\beta^\alpha (-k) \, ,
\label{n2}
\eea
which hold in any $\frep$.

The attention in \cite{Mintchev:2002zd, Mintchev:2003ue}
has been mainly focused on the subclass $\srep \subset \rep$ of
representations,
characterized by reflection matrices satisfying
\be
\cS_{12}(k_1,k_2)\R_2(k_1) = \R_2(k_1)\cS_{12}(-k_1,k_2)\, ,
\label{L}
\ee
which is stronger than the boundary Yang-Baxter equation (\ref{SRSR}).
We stress in this respect that $\cS$ and $\R$ in the impurity NLS model
obey (\ref{SRSR}) but not (\ref{L}), i.e. in our case $\frep \notin \srep$.

The boundary Yang--Baxter equation (\ref{SRSR}) is actually the
vacuum expectation
value of the defect exchange relation (\ref{rr}) in the representation $\frep$.
Taking the vacuum expectation value of the remaining relations
(\ref{tt},\ref{tr}),
one obtains the transmission Yang--Baxter equation
\bea
\cS_{12}(k_1 , k_2)\, \T_1(k_1)\,
\cS_{21}(k_2 , k_1)\, \T_2(k_2) =
\nonumber\\ {\T_2(k_2)}\, \cS_{12}(k_1 , k_2)\, \T_1(k_1) \,
\cS_{21}(k_2 , k_1)\, ,
\label{STST}
\eea
and the mixed reflection--transmission Yang--Baxter equation
\bea
\cS_{12}(k_1 , k_2)\, \T_1(k_1)\,
\cS_{21}(k_2 , k_1)\, \R_2(k_2) =
\nonumber\\
\R_2(k_2)\, \cS_{12}(k_1 , -k_2)\,
\T_1(k_1) \, \cS_{21} (-k_2 , k_1)\, .
\label{STSR}
\eea
The relations (\ref{STST},\ref{STSR}) have been discovered in
\cite{Mintchev:2003ue},
where it is shown that they are a general consequence of
(\ref{hanal}-\ref{unit},\ref{SRSR}).
The validity of (\ref{STST},\ref{STSR}) in our case can be checked directly,
inserting (\ref{rtmat},\ref{bulkS}).

At this stage we can define the basic structure $\{\Hil, \D, \Omega, \Phi\}$
in terms of $\frep$ as follows:

\begin{itemize}

\item {} $\Hil$, $\langle \cdot\, ,\, \cdot \rangle $ and $\Omega$ are
the Hilbert space, the scalar product and the vacuum state of
$\frep$, where $\{\R,\, \T\}$ and $\cS$ are given by
(\ref{coef1}-\ref{rtmat}) and (\ref{bulkS}) respectively.

\item {} The quantum fields $\Phi_\pm$, defined by (\ref{pm}),
admit the series representation (\ref{phiser}), where
\bea
\Phi^{(n)}_\pm (t,x) =
\int\prod_{i=1\atop j=0}^n
\frac{dp_i}{2\pi}\frac{dq_j}{2\pi}\,
a^{* \pm}(p_1)\ldots a^{* \pm}(p_n)a_\pm(q_n)\ldots
a_\pm(q_0) \cdot \nonumber \\
\frac{e^{i\sum\limits_{j=0}^n(q_j x-q^2_j t)-i
\sum\limits_{i=1}^n(p_i x-p^2_i t)}}
{\prod\limits_{i=1}^n(p_i-q_{i-1} \mp i\varepsilon)
(p_i-q_i \mp i\varepsilon)} \, .
\qquad \qquad \qquad
\label{phin}
\eea

\item {} The domain $\D$ is the finite particle subspace of $\frep$,
which is well--known to be dense in $\Hil$.

\end{itemize}

The mere fact that our system interacts with an impurity shows up at the
algebraic level, turning the Zamolodchikov--Faddeev (ZF) algebra from
the impurity--free case \cite{Sklyanin:ye}--\cite{Davies:gc}
to an RT algebra ({\ref{aa}--\ref{unit3}).
The details characterizing the impurity enter the construction at the
level of representation by means of the reflection and transmission matrices
(\ref{rtmat}). Notice also that the series (\ref{phiser}) is actually
a finite sum
when $\Phi$ is acting on $\D$. The coupling constant $g$ appears
explicitly in (\ref{phiser}) and implicitly in
$a_\alpha$ and $a^{\ast \alpha}$ which depend on $g$ through $\cS$.
The properties of the quantum field $\Phi$, defined above,
are summarized in the following

\bigskip
{\bf Proposition:} $\Phi (t,x)$ is a well--defined operator--valued
distribution
satisfying the canonical commutation relations
(\ref{ccr1},\ref{ccr2}) on $\D$, as well as
the equation of motion
\be
(i\prt_t + \prt_x^2 )\langle \varphi \, ,\, \Phi (t,x) \psi \rangle
= 2g\, \langle \varphi \, ,\, :\Phi \Phi^* \Phi: (t,x) \psi \rangle \, ,
\qquad x\not=0\, ,
\label{qeqm}
\ee
and the boundary conditions
\be
\lim_{x \downarrow 0}
\left(\begin{array}{cc} \langle \varphi \, ,\, \Phi (t,x) \psi \rangle  \\
\prt_x \langle \varphi \, ,\, \Phi (t,x) \psi \rangle
\end{array}\right) = \alpha
\left(\begin{array}{cc} a & b\\ c&d\end{array}\right)
\lim_{x \uparrow 0}
\left(\begin{array}{cc} \langle \varphi \, ,\, \Phi (t,x) \psi \rangle  \\
\prt_x \langle \varphi \, ,\, \Phi (t,x) \psi \rangle \end{array}\right) \, ,
\label{qbc}
\ee
\be
\lim_{x\to \pm \infty} \langle \varphi \, ,\, \Phi (t,x) \psi \rangle = 0 \, ,
\label{qbcinf}
\ee
for any $\varphi,\, \psi \in \D$.
\bigskip

{}For the proof of this statement we refer to \cite{CMR}, where the
$\delta$--impurity
(see eq. (\ref{deltaimp})) is considered
in detail. Following \cite{Gattobigio:1998si},
the normal product $:\cdots :$ in (\ref{qeqm}) preserves the
original order of the creators; the original order of two
annihilators is preserved if both belong to the same $\Phi$ or $\Phi^*$ and
inverted otherwise. Since $\Phi $ and the hermitian conjugate
$\Phi^*$ are unbounded operators,
the delicate points in proving the above proposition are essentially
domain problems.
They are solved taking into account that the reflection and
transmission amplitudes
$R_+^+$ and $T_+^-$ ($R_-^-$ and $T_-^+$) have
no poles in the complex upper (lower) half--plane, which is a consequence of
condition (\ref{nobs}) ensuring the absence of impurity bound states.

For $\alpha = a = d =1$ and $b = c = 0$ one expects to
recover from (\ref{phin}) the solution of the NLS equation without impurity.
We will show now that this is indeed the case. First of all we observe that
in this limit
\be
\R(k) = 0\, , \qquad
\T(k) =\left(\begin{array}{cc} 0 & 1\\1 & 0\end{array}\right)\, .
\label{noimp1}
\ee
Because of (\ref{RT1},\ref{RT2}), in the classical case one finds
$\lambda_-(k) = \lambda_+(k)$ and $\Phi $ defined by 
(\ref{pm}--\ref{class}) precisely
reproduces the classical solution without impurity.
The quantum case is slightly more involved. The data (\ref{noimp1}) fix a
Fock representation of $\alg$ in which
\be
r (k) = 0 \, , \qquad t(k) = \left(\begin{array}{cc} 0 & t^-_+\\t^+_- 
& 0\end{array}\right)\, .
\label{noimp2}
\ee
{}From eq. (\ref{unit2}) one deduces that
\be
t_+^-\, t_-^+ = t_-^+\, t_+^- = {\bf 1}\, ,
\label{noimp3}
\ee
where $\bf 1$ is the identity operator in $\Hil$. We stress however
that $t_+^-$ and $t_-^+$ are {\it not} proportional to $\bf 1$, since
they do not commute with $a_\pm(k)$ (see eq. (\ref{at})).
In agreement with this fact and consistently with
the exchange relations (\ref{aa}--\ref{aa*}) and the
form of the bulk scattering matrix, one has $a_-(k) \not=a_+(k)$.
Therefore the argument used at the classical level does not apply and
one must proceed in the quantum case differently. We observe in this
respect that inserting (\ref{noimp2}) in (\ref{aa*}),
one concludes that the polynomials of the operators
$\{a^{\ast +}(k),\, a_+(k),\, {\bf 1}\}$
close a ZF algebra $\A_+$ with exchange factor
$\cS_{+ + }^{+ +}(k_1 , k_2) = S(k_1-k_2)$.
Applied on the vacuum $\Omega$, the elements of $\A_+$ generate
a subspace $\Hil_+\subset \Hil$. By construction the quantum field 
$\Phi_+$ leaves
invariant $\D_+=\D \cap \Hil_+$ and its restriction 
$\Phi_+\vert_{\D_+}$ on $\D_+$
solves \cite{Davies:gc} the impurity-free NLS equation. Analogously,
the algebra $\A_-$ generated by $\{a^{\ast -}(k),\, a_-(k),\, {\bf 1}\}$
is a ZF algebra with exchange factor $\cS_{- - }^{- -}(k_1 , k_2) = 
S(-k_1+k_2)$.
The counterpart $\Hil_-$ of $\Hil_+$ defines the domain
$\D_-=\D \cap \Hil_-$, which is invariant under $\Phi_-$.
The restriction $\Phi_-\vert_{\D_-}$ is also a solution of the NLS equation
without impurity. Being related by a parity transformation $x \mapsto -x$,
which is a symmetry in this case,
$\Phi_+\vert_{\D_+}$ and $\Phi_-\vert_{\D_-}$
are unitary equivalent. Finally, the fact that in momentum
space parity is implemented by
$k \mapsto -k$, explains the relation
\be
\cS_{+ + }^{+ +}(k_1 , k_2) = \cS_{- - }^{- -}(-k_1 , -k_2) \, .
\ee

Turning back to the general impurity case, one can directly verify by 
means of (\ref{phin})
that the Hamiltonian $H$, which generates the
time  evolution
\be
\Phi(t,x) = \e^{itH}\, \Phi(0,x)\, \e^{-itH}\, ,
\label{tevolution}
\ee
has the familiar quadratic form
\be
H = \int \frac{dk}{2\pi} k^2 a^{\ast \alpha }(k) a_\alpha (k) \, .
\label{ham}
\ee
H is actually the second term of a whole sequence 
\cite{Mintchev:2003kh, Ragoucy:2004sw}
\be
H_n = \int \frac{dk}{2\pi} k^{2n} a^{\ast \alpha }(k) a_\alpha (k) \, 
, \qquad n = 0,1,2, ...
\label{intofm}
\ee
of integrals of motion in involution. In this sense the impurity system under
consideration is integrable. The simple form of $H_n$ is among the advantages
of the RT algebra approach.

Employing (\ref{phiser},\ref{phin}), one can construct all
correlation functions of
$\Phi$ and $\Phi^*$. The structure of (\ref{phin}) implies
that for the $2n$--point function one needs at most the
$(n-1)$--th order contribution in (\ref{phiser}). In fact, one has for example
\bea
\langle \Omega, \Phi(t_1,x_1)\Phi^\ast (t_2,x_2)\Omega \rangle =
\int_{-\infty}^{+\infty} \frac{dk}{2\pi} \e^{-ik^2t_{12}} \cdot
\qquad \qquad \qquad \nonumber \\
\Big\{\theta (x_1)\theta (x_2)\left [\e^{ikx_{12}} +
R_+^+(k)\e^{ik{\widetilde x}_{12}}\right ]+
\theta (-x_1)\theta (-x_2)\left [\e^{ikx_{12}} +
R_-^-(k)\e^{ik{\widetilde x}_{12}}\right] +
\nonumber \\
\theta (x_1)\theta (-x_2) T_+^-(k)\e^{ikx_{12}} +
\theta (-x_1)\theta (x_2) T_-^+(k)\e^{ikx_{12}}\Big\} \,  ,
\qquad \qquad \quad
\label{two-point}
\eea
where $t_{12} = t_1-t_2$, $x_{12} = x_1-x_2$ and ${\widetilde x}_{12}
= x_1+x_2$.
Analogous, but more involved integral representations
hold for the $2n$-point functions with $n>1$.

In \cite{CMR} it is also shown that $\Phi$ and $\Phi^*$ admit
asymptotic limits in a suitably
adapted to the impurity case Haag--Ruelle scattering theory. The net
result is that the asymptotic states are obtained applying the
creation operators $a^{*\pm}$ to the vacuum $\Omega$:
in $\RR_+$  and $\RR_-$ the asymptotic incoming particles are generated by
$\{a^{*+}(k)\, :\, k<0\}$ and $\{a^{*-}(k)\, :\, k>0\}$ respectively,
while the outgoing particles are created by $\{a^{*+}(k)\, :\,
k>0\}$ and $\{a^{*-}(k)\, :\, k<0\}$.
The scattering amplitudes are thus derived in a purely
algebraic way, using the exchange relation (\ref{aa*}) and the fact that
according to (\ref{annihilate}) $a_\pm$ annihilate $\Omega$.
As expected, the total scattering operator $\bf S$ factorizes, the factors
being the bulk scattering matrix $\cS$ (\ref{bulkS}) and the reflection and
transition matrices $\R$ and $\T$ (\ref{rtmat}).

Summarising, we have established a family (\ref{bc}-\ref{nobs}) of
point--like impurities
interacting with the NLS field, which preserve quantum integrability.
These systems can be investigated by the inverse scattering method.
We have shown
in this respect that the RT algebra $\alg$ and its Fock representation
$\frep$ allow to construct not only the scattering operator but also the
off--shell quantum field $\Phi(t,x)$.

\sect{Discussion}

A debated and physically relevant question in the theory of integrable
systems with impurities concerns the space--time symmetry of the {\it
bulk} scattering
matrix $\cS$. It is well--known that impurities break down Galilean
(Lorentz) invariance in the {\it total} scattering matrix ${\bf S}$.
However, since $\cS$ describes the scattering away from the impurity,
one might be tempted
to assume \cite{Delfino:1994nr}--\cite{Castro-Alvaredo:2002dj}
that $\cS$ preserves these symmetries and that the breaking in
${\bf S}$ is generated exclusively by the reflection and
transmission coefficients $\R$ and $\T$. Unfortunately however, the
conditions of factorized scattering then imply
\cite{Delfino:1994nr, Castro-Alvaredo:2002dj} that
$\cS$ is constant. Being too restrictive, this property limits very much the
interest in such systems. In order to avoid the problem,
a consistent factorized scattering theory of a unitary scattering operator
has been developed in \cite{Mintchev:2002zd, Mintchev:2003ue} in terms of
RT algebras, without necessarily assuming that $\cS$ is Galilean
(Lorentz) invariant.
The impurity NLS model considered above, is the first concrete application
of this framework with non--trivial bulk scattering. The lesson from
it is quite
instructive. Focusing on (\ref{bulkS}), we see that Galilean
invariance is indeed broken by the entries of $\cS$, which
describe the scattering of two incoming particles
localized for $t\to - \infty$ on $\RR_-$ and $\RR_+$ respectively. In fact,
these entries depend on $k_1+k_2$ and not on $k_1-k_2$. The intuitive
reason behind this breaking is that before such particles scatter, one of them
must necessarily cross the impurity. The non--trivial transmission is
therefore the origin of the symmetry breaking in $\cS$.
This conclusion agrees with the observation that in systems
which allow only reflection (e.g. models on the half--line), one can have
both Galilean (Lorentz) invariant and non--constant bulk scattering matrices.

{}For simplicity we focused in this paper on linear impurity boundary
conditions. One can expect however that there exist non-liner
boundary conditions of the type proposed in \cite{Bowcock:2003dr, 
Bowcock:2004my}
for the Toda model, which also preserve the integrability of the NLS equation.

Another aspect which deserves further investigation is the issue of internal
symmetries in the presence of impurities. This question has been
partially addressed in \cite{Mintchev:2003kh, Ragoucy:2004sw}, where the role
of the reflection and transmission elements of the RT algebra as
symmetry generators
has been established. It will be interesting in this respect
to extend the analysis \cite{Mintchev:2001aq} of the $SU(N)$--NLS model on
the half--line to the impurity case.

Let us conclude by observing that the concept of RT algebra indeed represents
a powerful tool for solving the NLS model with impurities.
We strongly believe that this algebraic framework is actually universal
and applies to the quantization of other systems as well.

\bigskip

{\bf Acknowledgments:} We thank the referee for the constructive criticism.

\end{document}